\documentclass[aps,prl,reprint]{revtex4-1}
\usepackage{graphicx,amsmath,amssymb,amsfonts,latexsym,color,dcolumn,bm,mathtools}
\usepackage{verbatim}
\usepackage{pdfpages} 
\usepackage{textcomp}

\begin{document}
\title{Overwhelming thermomechanical motion with microwave radiation pressure shot noise}
\author{J. D. Teufel} 
\email[To whom all correspondence should be addressed. ]{john.teufel@nist.gov}
\affiliation{National Institute of Standards and Technology, 325 Broadway, Boulder, CO 80305, USA}
\author{F. Lecocq}
\affiliation{National Institute of Standards and Technology, 325 Broadway, Boulder, CO 80305, USA}
\author{R. W. Simmonds}
\affiliation{National Institute of Standards and Technology, 325 Broadway, Boulder, CO 80305, USA}

\begin{abstract}
We measure the fundamental noise processes associated with a continuous linear position measurement of a micromechanical membrane incorporated in a microwave cavity optomechanical circuit. We observe the trade-off between the two fundamental sources of noises that enforce the standard quantum limit: the measurement imprecision and radiation-pressure backaction from photon shot noise.  We demonstrate that the quantum backaction of the measurement can overwhelm the intrinsic thermal motion by 24~dB, entering a new regime for cavity optomechanical systems.  
\end{abstract} 
\maketitle
Quantum mechanics places limits on the precision with which one can simultaneously measure canonically conjugate variables, such as the position and momentum of an object.  For example, if one scatters light off an object to determine its position, the momentum kicks from individual photons will necessarily apply forces back on that object. In the limit of a continuous measurement in which a coherent state of light is used to infer the state of the mechanical degree of freedom, it is the quantum statistics of the photons that will place simultaneous limits on the information gained and the backaction imparted. So while a stronger measurement reduces the imprecision, this comes at the cost of increasing radiation pressure shot noise forces. The minimum noise added by the measurement process is bounded by the standard quantum limit (SQL) and occurs when the imprecision noise and backaction noise are perfectly balanced\cite{Caves1980,Clerk2010}. Experimentally observing this fundamental trade-off requires a system that interacts strongly with the measurement yet weakly with its thermal environment. 

\begin{figure} 
\includegraphics[width=86mm]{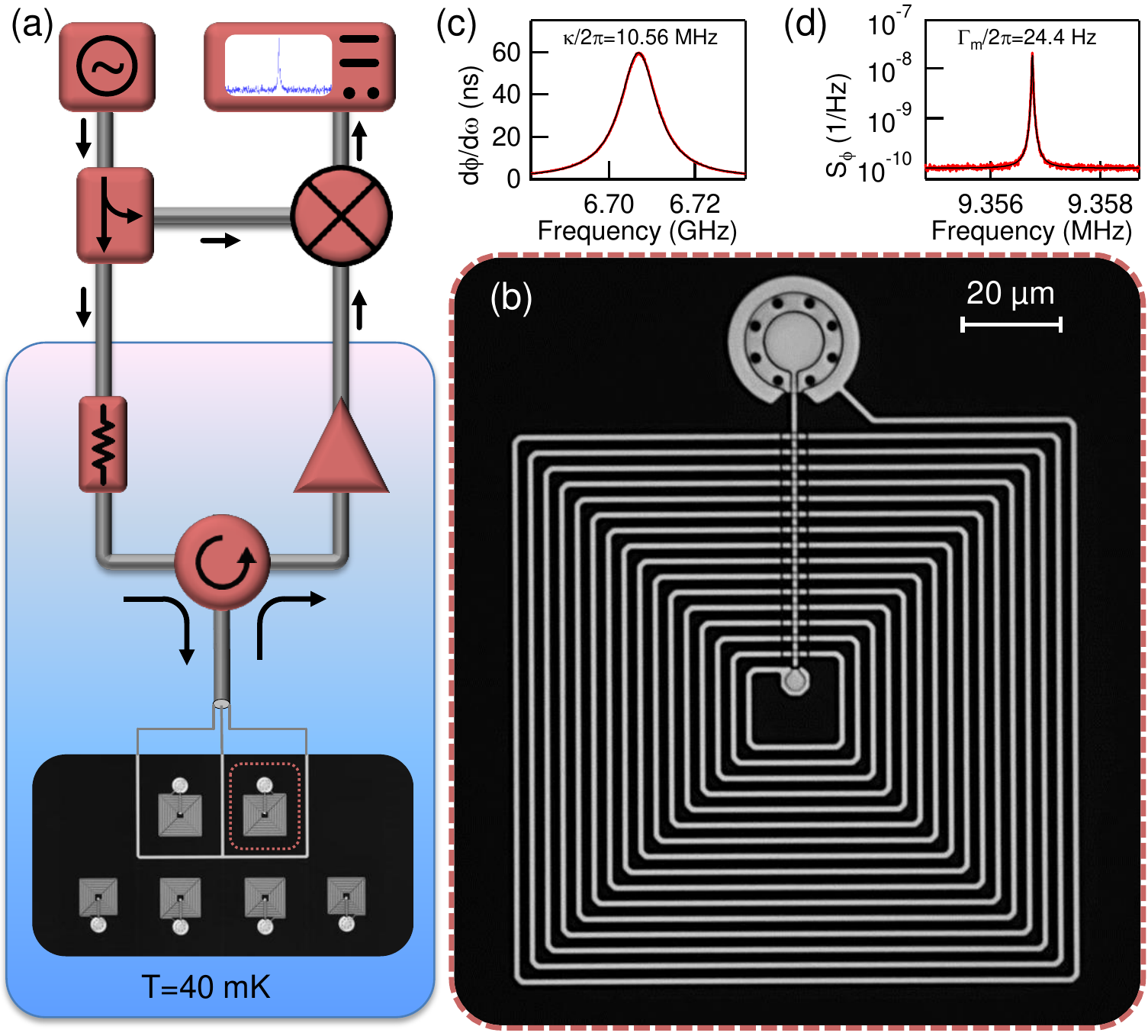}
\caption{\textbf{Experimental schematic and device characterization.}
\textbf{(a)} We interrogate an array of microwave cavity optomechanical circuits with a cryogenic interferometer.  A coherent microwave drive at the cavity resonance frequency is inductively coupled to the resonators.  The reflected microwave field acquires phase modulation sidebands from the mechanical mode that are amplified cryogenically and demodulated at room temperature.  
\textbf{(b)}  Optical micrograph of a single microwave cavity optomechanical circuit, microfabricated out of aluminum (grey) on a sapphire substrate (black).  The mechanically compliant capacitor resonates with a spiral inductor to form a superconducting LC resonant circuit.  
\textbf{(c)}  A Lorentzian fit (black) to the microwave group delay (red) shows a highly overcoupled electromagnetic resonance at $\omega_\mathrm{c}/2\pi=
6.707$~GHz with a total linewidth $\kappa/2\pi=10.56$~MHz.
\textbf{(d)}  When this microwave resonance is driven at $\omega_\mathrm{c}$,
the noise spectrum of the microwave phase fluctuations shows a mechanical resonance (red) whose fit (black) yields $\Omega_\mathrm{m}/2\pi=9.357$~MHz 
with intrinsic linewidth of $\Gamma_\mathrm{m}/2\pi=24.4$~Hz.}
\label{fig1}
\end{figure}

One successful strategy for the measurement of mechanical systems has been that of cavity optomechanics\cite{Aspelmeyer2014}, in which the displacement of a mechanical oscillator of frequency $\Omega_\mathrm{m}$ tunes the resonance frequency $\omega_\mathrm{c}$ of a high-frequency electromagnetic cavity.  When the cavity is excited with a coherent drive of power $P$ at $\omega_\mathrm{c}$, the mechanical motion becomes encoded as phase fluctuations of the reflected light. As the measurement power is increased this mechanical signal rises above the measurement noise floor, ideally limited by the vacuum fluctuations in the phase quadrature of the drive.  This improvement in the measurement imprecision comes at the expense of an increasing backaction force, ideally solely from the vacuum fluctuations in the amplitude quadrature of the drive.  

The measured total displacement spectral density $S_x$ at the mechanical resonance frequency is a combination of the zero-point motion $S_x^\mathrm{zp}$, the thermal motion $S_x^\mathrm{th}$, the measurement imprecision $S_x^\mathrm{imp}$ and measurement backaction $S_x^\mathrm{ba}$. The SQL can be stated by the inequality $S_x^\mathrm{imp}+S_x^\mathrm{ba}\ge S_x^\mathrm{zp}$. For an ideal homodyne detection of the phase quadrature of the light, the added noise of the measurement at $\Omega_\mathrm{m}$ is
\begin{equation}\label{eq1}
S_x^\mathrm{imp}+S_x^\mathrm{ba}=\frac{S_x^\mathrm{zp}}{2}\left(\frac{P_\mathrm{SQL}}{P}+\frac{P}{P_\mathrm{SQL}}\right).
\end{equation}
The power required to reach the SQL for an ideal, lossless cavity optomechanical system is $P_\mathrm{SQL}=\hbar\omega_\mathrm{c}\Gamma_\mathrm{m}\left(\kappa^2+4\Omega_\mathrm{m}^2\right)/64g_0^2$, where $\Gamma_\mathrm{m}$ and $\kappa$ are the mechanical and cavity dissipation rates, respectively, and $g_0$ is the vacuum optomechanical coupling rate\cite{Clerk2010,Aspelmeyer2014}. It is only at the optimum power $P_\mathrm{SQL}$ that the measurement is impedance matched to the mechanical system\cite{Clerk2010}, reaching the minimum added noise of $S_x^\mathrm{zp}=2\hbar/\left(m\Omega_\mathrm{m}\Gamma_\mathrm{m}\right)$, where $m$ is the mass of the oscillator. 
 
In practice, however, any technical deficiencies, such as internal loss of the cavity, inefficient homodyne detection or excess noise on the coherent drive would preclude reaching this fundamental limit.  While many experiments are now able to engineer low enough $P_\mathrm{SQL}$ to be achieved in the laboratory, the quantum backaction from the radiation pressure shot noise is often obscured by residual thermal motion of the oscillator \cite{Teufel2009, Anetsberger2010,Westphal2012}. It is only recently that cavity optomechanical experiments observe backaction noise on par with the thermal occupancy\cite{Purdy2013a,Schreppler2014,Suh2014,Wilson2014}. While clever cross-correlation techniques can separate the thermal motion from the effects of the measurement\cite{Heidmann1997,Borkje2010,Purdy2013a}, one would ideally like to operate at much larger powers where the quantum backaction dominates all other sources of noise including the thermal motion, $S_x^\mathrm{ba}\gg S_x^\mathrm{th}$. It is in this high power limit where many interesting quantum phenomena reach their full potential; these effects include, for example, ponderomotive squeezing\cite{Fabre1994} or amplification\cite{Botter2012} of light and coherent feedback cooling of motion\cite{Mancini1998}.

\begin{figure} 
\includegraphics[width=86mm]{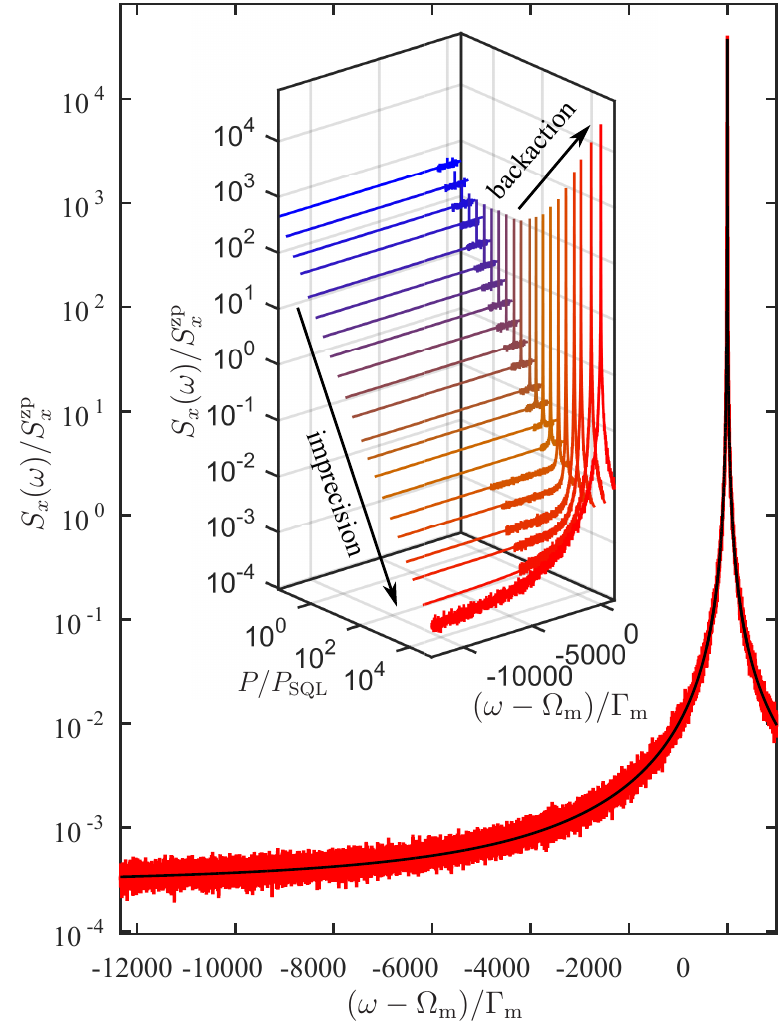}
\caption{\textbf{Calibrated mechanical noise spectra.}
Phase noise spectra in units of normalized displacement spectral density allow for direct noise thermometry of the occupancy of the mechanical mode for various measurement strengths.  Even at weak microwave drive powers ($\approx$$10$~fW), we readily resolve the thermomechanical motion of the oscillator above measurement imprecision with negligible backaction.  As the drive power is increased, the imprecision noise floor is lowered, and radiation pressure shot noise of the microwave photons imparts quantum backaction on the mechanical mode as visible from the increased height of the Lorentzian peak. At the largest measurement power, the quantum backaction completely overwhelms the intrinsic thermal motion, and the backaction exceeds the measurement imprecision over a frequency span of more than $10^4$ mechanical linewidths. For all of the data shown here, the measured mechanical linewidth remains within $5\%$ of the intrinsic value $\Gamma_\mathrm{m}/2\pi=24.4$~Hz.  For the highest measurement power $P=7.8$~nW, this condition corresponds to residual detuning of less than 1 part per million of $\kappa$.
} 
\label{fig2} 
\end{figure}

We achieve this limit in a microwave cavity optomechanical circuit incorporated in a cryogenic interferometer, as shown in Fig.\ref{fig1}. The circuit is made out of aluminum and  consists of a superconducting $15$~nH spiral inductor resonating with a vacuum-gap, parallel-plate capacitor\cite{Cicak2010}. This LC resonant circuit creates a microwave ``cavity" whose resonance frequency depends sensitively on the separation between the capacitor electrodes. The mean plate separation of $40$~nm yields a cavity frequency of $\omega_\mathrm{c}/2\pi=6.707$~GHz. We probe the cavity with a coherent microwave drive applied at $\omega_\mathrm{c}$, heavily filtered and  attenuated at cryogenic temperatures to ensure that the microwave field is in a pure coherent state.  The reflected signal is separated from the incident wave through a microwave circulator, amplified with a cryogenic low-noise amplifier, and the phase quadrature is measured with a homodyne detection scheme. As the cavity is highly overcoupled to its feed-line, very few photons are dissipated at the device, and the cavity resonance is best characterized by the Lorentzian dependence of the group delay as a function of frequency (Fig.~\ref{fig1}c). With internal losses contributing to less than $1$\% to the total linewidth $\kappa/2\pi=10.56$~MHz, the cavity photons are efficiently collected by the measurement apparatus, and the cavity is devoid of thermal excitations.

The top plate of the capacitor is mechanically compliant and has a fundamental flexural mode at $\Omega_\mathrm{m}/2\pi=9.357$~MHz, which strongly couples to the microwave resonance \cite{Teufel2011a}. With a total mass of $85$~pg, it has a zero-point motion of  $x_\mathrm{zp}=\sqrt{\hbar/(2m\Omega_\mathrm{m})}=3.3$~fm. At a cryostat temperature of $40$~mK the mechanical mode has an equilibrium occupancy of $n_\mathrm{th}=[\exp(\hbar\Omega_{\mathrm{m}}/k_{\mathrm{B}}T)-1]^{-1}=90$ phonons. When the overcoupled optomechanical cavity is driven exactly on resonance, the thermomechanical signal appears only in the phase quadrature of the microwave light, and the measured power spectral density of the phase quadrature of the microwave field $S_\phi$ is directly related to the displacement spectral density $S_x$ via the relation
\begin{equation}
\label{eq2}
\frac{S_x}{x_\mathrm{zp}^2}= \left(\frac{\kappa^2+4\Omega_\mathrm{m}^2}{64g_0^2}\right) S_\phi.   
\end{equation}

In Fig.\ref{fig1}d the mechanical line shape is evident from the Lorentzian peak in the power spectral density of the phase fluctuations $S_\phi$ centered at $\Omega_\mathrm{m}$, yielding a intrinsic mechanical damping rate of $\Gamma_{\mathrm{m}}/2\pi=24.4$~Hz. As with previous experiments \cite{Regal2008,Teufel2009,Suh2014}, we use the thermal motion in the weak measurement regime to calibrate this transduction and determine $g_0/2\pi =230$~Hz. This completely determines $P_\mathrm{SQL}=91.4$~fW. With these parameters determined, we may explore the power dependence of the calibrated noise spectra. 

\begin{figure}
\includegraphics[width=86mm]{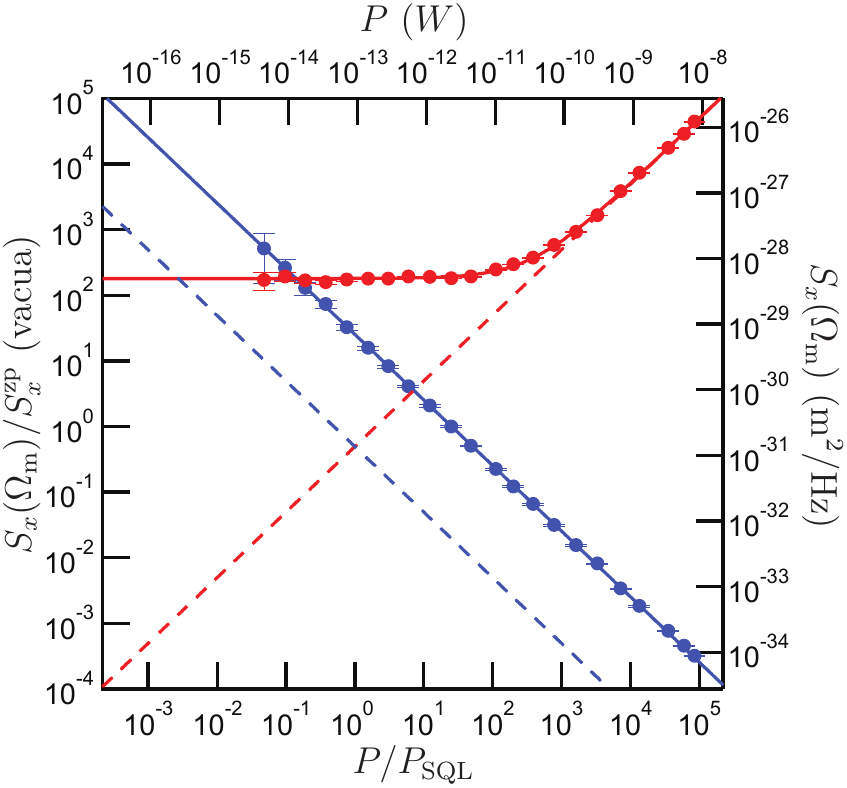}
\caption{\textbf{Fundamental trade-off between imprecision and measurement backaction.}
The total noise power at the mechanical resonance frequency is decomposed into its two contributions. The data show the actual motion (red circles) and the apparent motion (blue circles) as a function of measurement strength. The dashed lines show the expected imprecision and backaction of the perfect quantum measurement.  The solid lines are  theoretical predictions for the imprecision (blue) and total motion (red) based on the independently determined measurement efficiency and thermal decoherence rate.  At the highest power, the imprecision is lowered to $8.8\times 10^{-35}$~m$^{2}$/Hz, a factor of 1700 below that at the standard quantum limit. Likewise, the quantum backaction from momentum kicks of the measurement photons overwhelms the intrinsic thermal motion of $n_\mathrm{th}=90$, demonstrating a quantum measurement rate 250 times larger than the thermal decoherence rate.}
\label{fig3}
\end{figure}

As shown in Fig.~\ref{fig2}, increasing the measurement power has two effects on the measured displacement spectral density.  First, the measurement imprecision decreases with increasing power as expected. For moderate measurement strengths ($P\lesssim 100$~fW), this reduced imprecision is the only observable effect as the expected backaction remains well below the thermal motion.  Beyond this power, the observed height of the Lorentzian peak begins to grow linearly with increasing drive power.  It is in this regime where we can begin to observe the fundamental trade-off expected from Eq.~\ref{eq1}. From Lorentzian fits to these noise spectra, we independently determine the imprecision noise and actual motion of the oscillator in the presence of backaction.  Figure~\ref{fig3} shows these two components of the noise as a function of the measurement strength. The imprecision follows the expected $1/P$ dependence with no visible deviation even at the highest power.  The solid blue line is the predicted imprecision. It is degraded from the ideal imprecision (blue dashed line) by the independently determined measurement homodyne efficiency $\eta=0.02$. This inefficiency derives primarily from the added noise of the microwave amplifier operating at $4$ kelvin and could be improved by at least an order of magnitude with the incorporation of a superconducting parametric amplifier\cite{Teufel2009,Teufel2011b}. The red dashed line is the expected quantum backaction of the measurement. The quantitative agreement between the expected backaction and measured data confirms that there is no excess noise in the cavity or the drive, and demonstrates a true coherent state of the microwave field. At the optimum measurement power of $P=P_\mathrm{SQL}/\sqrt{\eta}\approx 600$~fW, the total added noise is optimized, yielding $S_x^\mathrm{imp}\approx S_x^\mathrm{ba}\approx 3.6\times S_x^\mathrm{zp}$. It is at this power that the measurement is most sensitive to both the thermal mechanical environment and external forces, demonstrating a force sensitivity of $5.5$~aN/$\sqrt{\mathrm{Hz}}$ at the mechanical resonance frequency. 

At the highest measurement power of $P=7.8$~nW, a power approaching  $\approx 10^5\times P_\mathrm{SQL}$, we interrogate the system with approximately $10^{15}$ photons per second. Despite a modest detection efficiency, we achieve an imprecision noise of $8.8\times 10^{-35}$~m$^{2}$/Hz, $35$~dB below the zero-point level $S_x^\mathrm{zp}$. Concurrently, the quantum backaction increases the total mechanical motion to $2.2\times 10^{4}$ quanta. Comparing this to the intrinsic thermal occupancy of the mechanical mode, we observe microwave radiation pressure shot noise $24$~dB above the thermal motion.  In other words, the mechanical oscillator is so strongly coupled to the measurement that its dissipative thermal environment only contributes less than $0.4\%$ of its total motion.

These results demonstrate the fundamental sensitivity limits for mechanical measurements, with numerous applications \cite{Metcalfe2014} including areas such as magnetic resonance force microscopy\cite{Rugar2004} and the detection of gravitational waves\cite{Adhikari2014,Schnabel2010}. While future improvements such as enhanced detection efficiency would help minimize the total displacement noise,  the regime of strong measurement backaction demonstrated here opens the door toward an array of experiments utilizing quantum states of both light and motion. In order to reduce the excess thermal motion of the oscillator, one could implement active feedback to cool the motion close to the ground state\cite{Wilson2014,Krause2015}. Beyond simply using the light to interrogate the state of the mechanical oscillator, it is in this regime that an optomechanical system can prepare and measure nonclassical states of the light field.  One could achieve strong ponderomotive squeezing of the outgoing light \cite{Safavi2013,Purdy2013b}, and the parameters demonstrated here would predict ponderomotive squeezing of microwave light more than $20$~dB below vacuum. Likewise, the same correlations that can squeeze the microwave field can also be used to amplify the light in a nearly ideal way \cite{Botter2012,Metelmann2014}.  Finally, it is in this regime that an optomechanical system realizes a quantum nondemolition measurement of light\cite{Jacobs1994}.


\bibliography{Backaction_Bib}

\end{document}